\documentclass{article}

\usepackage[american]{babel}

\usepackage[letterpaper,top=2cm,bottom=2cm,left=3cm,right=3cm,marginparwidth=1.75cm]{geometry}

\usepackage{amsmath}
\usepackage{graphicx}

\usepackage{booktabs}                 
\usepackage{tabularx}                 
\usepackage{array}
\usepackage{ragged2e}
\usepackage{caption}
\usepackage[most]{tcolorbox}                 
\newcolumntype{Y}{>{\RaggedRight\arraybackslash}X}
\newtcolorbox{illusbox}{%
  enhanced, sharp corners=all, colback=white, colframe=black!45,
  boxrule=0.5pt, left=8pt, right=8pt, top=8pt, bottom=8pt,
  breakable=false, before skip=8pt, after skip=8pt}

\usepackage{csquotes}
\usepackage[style=apa,backend=biber]{biblatex}
\DeclareLanguageMapping{american}{american-apa}
\usepackage[colorlinks=true, allcolors=blue]{hyperref}

\title{\textbf{AI-Integrated Scientific Inquiry:\\ A Practice-Centered Vision for Science Education}}
\author{Arne Bewersdorff, Matias Rojas, and Xiaoming Zhai\\[0.6em]
  {\normalsize AI4STEM Education Center, University of Georgia, Athens, GA, USA}\\[0.3em]
  {\normalsize Corresponding author: Arne Bewersdorff, \texttt{arne.bewersdorff@uga.edu}}}
\date{}

\begin{document}
\maketitle

\begin{abstract}
\noindent Artificial intelligence (AI) has become part of scientific inquiry. Scientists use AI to observe and measure phenomena, to identify patterns in data, and to build models. As AI moves into scientific inquiry, it gains relevance for science education: students should learn how AI is changing scientific practices, ideally by engaging in AI-integrated scientific inquiry themselves. How to design such instruction, grounded in authentic scientific practice rather than taught as a standalone topic, remains an open question. In our vision, which we describe in this article, AI is treated as a set of scientific instruments that students use within the scientific practices described by the Next Generation Science Standards. Each instrument is a genuine scientific tool, pedagogically bounded: its controls are simplified while its core scientific function is preserved. The approach has two aims: engaging students in authentic scientific inquiry, and building an understanding of how AI is used in science and where it can mislead (discipline-based AI literacy, DAIL). In the article, we focus on the investigative core of inquiry, namely observing, analyzing, and modeling, and describe one exemplary AI instrument for each: computer vision for observing, clustering for analyzing, and generative modeling for modeling. We argue that every AI instrument in science education should carry a distinct reflection point that prompts critical evaluation of the AI instrument itself. Finally, we describe how agentic AI, operating across the whole inquiry rather than a single practice, could be represented, arguing that students should first build a foundational understanding of scientific inquiry and AI instruments before relying on agentic AI.

\vspace{0.8em}
\end{abstract}

\section{Introduction}
\label{sec:introduction}

Artificial intelligence (AI) has moved toward the center of scientific work. AI now supports researchers across the natural sciences as they generate hypotheses, design and conduct experiments, and extract meaning from data at scales beyond the reach of any single research team \parencite{jordan2015,wang2023}. This development runs across methods and disciplines: deep neural networks identify and count wild animals in millions of camera-trap photographs \parencite{norouzzadeh2018}. Deep learning predicts three-dimensional protein structures from sequence \parencite{jumper2021}, and generative models design entirely new proteins that do not occur in nature \parencite{watson2023}. Unsupervised methods cluster astronomical and genomic data and reveal patterns that were not anticipated \parencite{wang2023}. In each of these cases, AI does not replace the scientist; it functions as a new kind of instrument, one that must still be aimed, calibrated, and reflected upon.

As AI becomes part of scientific practices, there is a strong argument that it should become part of scientific literacy and of how science is learned \parencite{zhai2025}. For science teachers, the central question is not whether AI belongs in the classroom, but how to introduce it so that it strengthens and complements the authentic scientific practices students are already learning, rather than bypassing them by fostering AI literacy in isolation.

General AI literacy frameworks describe what every student should understand about AI. \textcite{long2020}, for example, describe the competencies a citizen needs to recognize and evaluate AI. These frameworks are valuable, but they are not written for the science teacher who wants to teach students how AI actually changes scientific practices. Recent work has begun to identify what is missing by distinguishing generic AI literacy from discipline-based (also called domain-specific) AI literacy (DAIL), the competencies required to use and evaluate AI within a particular field or discipline \parencite{knoth2024,zhai2025}. Related work has begun to map how AI is reshaping contemporary scientific practice and what this implies for science education \parencite{nehm2026}, and a first classroom step has introduced AI-based scientific inquiry using a machine-learning tool \parencite{herdliska2024}. What remains to be articulated is a concrete account of discipline-based AI instruction for science, one anchored in the actual practices and examples of scientific inquiry. We describe a vision and a platform that enacts it, places AI instruments across scientific inquiry, ties each to an authentic scientific practice, and ensures that students reflect on and critically evaluate the output of AI instruments. The platform is a browser-based learning environment that provides one reduced AI instrument for each scientific practice, with its reflection point built in; three of its instruments are shown in Figures~\ref{fig:cv}--\ref{fig:genmodel}. In this sense, we define an AI instrument as a tool that performs an act of scientific investigation that would ordinarily require human-like intelligence, such as finding patterns, inferring relationships, or generating structures. Such AI instruments lie along a continuum, from statistical learning such as regression and clustering, via deep learned models such as those used in computer vision, to generative AI. The boundary between statistics and machine learning is fluid \parencite{breiman2001}. Regression, though rooted in classical statistics, is also the usual entry point to supervised learning \parencite{hastie2009,russell2021}. What the instruments share is not a common mechanism but the critical evaluation they demand of the student. AI instruments differ, too, in how far their inner workings can be inspected, a point we return to in Section~\ref{subsec:operating}.

Beyond these AI instruments, more autonomous AI systems, including agentic AI, are becoming more than tools: they can assist researchers across multiple stages of scientific inquiry. The implications of this overarching role, often framed as AI as collaborator, are discussed in a separate subsection (Section~\ref{subsec:agentic}) and in the discussion that follows.

\section{A Vision for AI-Integrated Scientific Inquiry}
\label{sec:vision}

AI is not a new topic to be added ``on top'' of science education, but a new generation of scientific instruments, and these instruments belong within the scientific practices. The microscope did not turn biology into the study of lenses; it gave biologists a new way to observe, and with it a new question to ask: whether an observed structure is real or an artifact of the preparation. Computer vision, clustering, and generative models can be understood in the same way. The phenomenon under investigation remains what it has always been, such as why leaves differ in shape, how an object falls, or what makes a molecule sweet. AI serves as the instrument rather than the main object of investigation. At the same time, the role of AI in scientific work is something students are meant to learn about.

The two aims map onto scientific literacy (engaging in authentic scientific practices) and discipline-based AI literacy (understanding the AI instruments themselves). Using AI as instruments within scientific practices advances both together. The two aims are not separate lessons: students come to understand an AI instrument by using a pedagogically bounded version that preserves its core scientific function. These remain design claims about how such instruction should work; whether operating the instruments in fact produces the intended learning is an empirical question we return to (Section~\ref{subsec:challenges}).

\subsection{Scientific Practices as the Organizing Framework}
\label{subsec:practices}

To keep scientific inquiry at the center, the instruction is organized around the science and engineering practices \parencite{nrc2012} and operationalized in the Next Generation Science Standards \parencite{ngss2013}. The framework's eight practices range from asking questions, through planning and carrying out investigations and analyzing and interpreting data, to developing and using models, constructing explanations, and arguing from evidence. They provide a shared, standards-aligned structure onto which AI instruments can be mapped. Anchoring AI in scientific practices keeps the technology in the service of scientific inquiry: each AI instrument is placed at the practice it genuinely supports, and the practice itself, whether observation, analysis, or modeling, remains what the student is engaging in.

\subsection{Operating Real Instruments}
\label{subsec:operating}

A defining feature of the approach is that students operate genuine AI instruments that are pedagogically bounded and aligned, e.g.\ via reduced complexity, rather than reading about or abstractly discussing AI in science. An AI instrument in this sense is something the student aims, calibrates, and must understand in its essentials, such as a clustering algorithm that the student runs and then critically evaluates. Operating the instrument constitutes authentic scientific practice; understanding how it behaves and where it fails constitutes discipline-based AI literacy. This understanding is primarily functional rather than mechanistic: mechanistic understanding concerns how an instrument works internally, whereas functional understanding concerns what it does, when its output can be trusted, and how it can fail \parencite{hmelosilver2004,machamer2000}. A consistent boundary runs through every instrument: it computes, but the student interprets. The instrument reports the structure it finds, yet what that structure means, and whether it can be trusted, is the student's to decide. Unlike the microscope, whose workings a student can in principle inspect, AI instruments are often evaluated by their outputs without full mechanistic transparency. For a generative AI model, functional evaluation may be the primary form of evaluation available to the student, similar to how scientists often evaluate an opaque instrument through its outputs. A more transparent instrument such as clustering can in comparison be inspected more directly. Either way, the student judges the outcome the AI instrument produces.

The approach rests on two design commitments. First, each AI instrument is, while being a genuine scientific tool, pedagogically bounded: its interface and controls are simplified, while its core scientific function and the need for student interpretation are preserved. Second, each instrument is taught together with a characteristic reflection point---a consequential assumption, decision, or failure mode that prompts critical evaluation. Because an instrument works on the data it is given, reflection may concern the suitability of those data as well as the output itself. For clustering, for example, are the groups supported by the data or shaped by the variables included and the number of groups requested? For regression, does the selected function describe the data, or force a relationship through scattered points? For a generative model, did it produce what was specified, and is the output scientifically plausible? In the platform, this commitment is explicitly built into the AI instrument itself: an AI instrument should not be used for a scientific practice without first considering its potential reflection points. This is not a new form of critical thinking; it is the inherent skepticism of science, the same attention to controls, replication, and limits of traditional scientific instruments, directed at a new class of instruments. That continuity is what keeps the AI instrument within science education, with AI literacy embedded in the discipline, rather than constituting a separate unit on AI.

\section{AI Instruments Across Scientific Inquiry}
\label{sec:across}

This article focuses on the investigative core of scientific inquiry: observing, analyzing, and modeling. For each of these, we describe one AI instrument in some detail and identify related instruments more briefly, noting how AI is used for that practice in current research.

\begin{table}[htbp]
\centering
\caption{Three Illustrative AI Instruments Within Scientific Inquiry, Each Anchored to an NGSS Practice, an Authentic Scientific Example, and an Exemplary Reflection Point}
\label{tab:instruments}
\footnotesize
\begin{tabularx}{\textwidth}{@{}YYYY@{}}
\toprule
\textbf{Scientific practice (NGSS)} & \textbf{AI instruments} & \textbf{Authentic scientific example} & \textbf{Exemplary reflection point} \\
\midrule
Planning and Carrying Out Investigations (focus: observing) &
Computer vision: object detection \& counting; object tracking; audio classification &
Counting animals in camera-trap images \parencite{norouzzadeh2018} &
Did it count the intended objects, and how did characteristics of the images affect false detections and missed objects? \\
\addlinespace
Analyzing \& Interpreting Data &
Clustering; regression &
Finding structure in astronomical and genomic data \parencite{jordan2015} &
Are the groups supported by the data, or shaped by the variables included and the value of $k$? \\
\addlinespace
Developing \& Using Models &
Generative modeling; symbolic regression; ML surrogates &
Designing novel proteins and molecules from a specification \parencite{watson2023} &
Did it build what was specified, and do scientific constraints and available evidence support the intended property? \\
\bottomrule
\end{tabularx}
\end{table}

\subsection{Observing: Turning the World into Data}
\label{subsec:observing}

\begin{figure}[htbp]
\centering
\begin{illusbox}
{\centering\includegraphics[width=\linewidth]{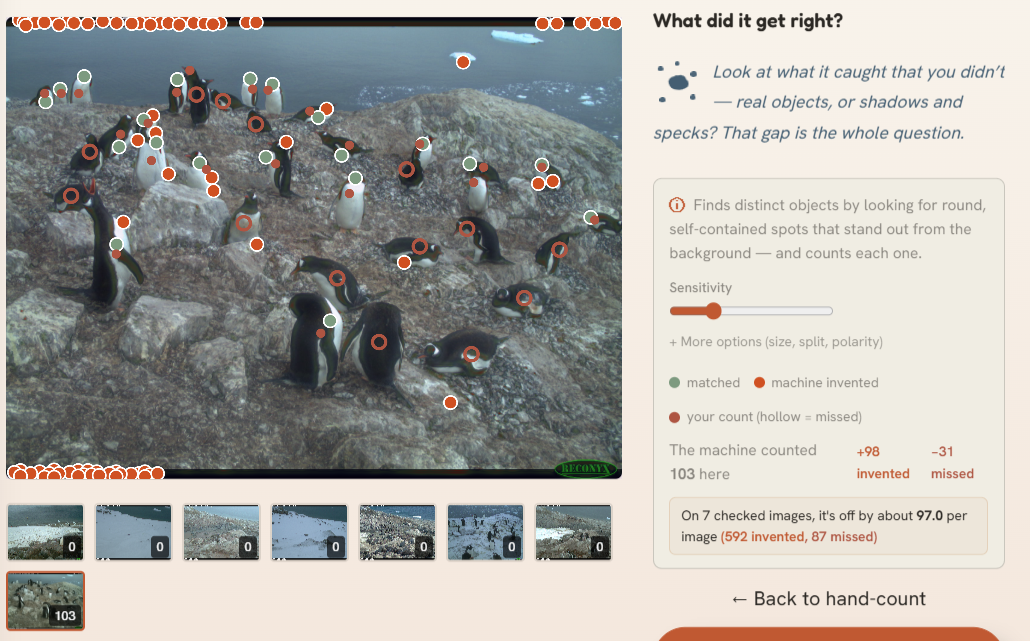}\par}
\captionof{figure}{Observing: Computer-vision counter}\label{fig:cv}
{\small\RaggedRight
\textbf{What it does:} Detects and counts objects in an image; the student picks a detection method and tunes its sensitivity.\par\smallskip
\textbf{Student's task:} Choose a method (edges, color, blob\ldots), set the threshold, and compare the instrument's count with a hand count of the same image.\par\smallskip
\textbf{Built-in check:} Overview of matched, falsely detected, and missed objects. The tool prompts students to consider how characteristics of the input image affected the count.\par\smallskip
\textbf{Across domains:} The same approach can be used to count animals in wildlife photographs, cells in a micrograph, or craters in a surface image, although defining the target and interpreting the count remain domain-specific.\par}
\end{illusbox}
\end{figure}

In ecology, computer vision detects and counts animals in aerial and camera-trap imagery, a task that previously required years of manual effort. The computational methods range from classical computer-vision detectors that threshold an image and count the resulting blobs \parencite{chabot2016} to deep neural networks that identify, count, and describe animals across millions of images \parencite{norouzzadeh2018}. Our first AI instrument is a pedagogically bounded version of this approach: students count objects across images, adjust the sensitivity, and compare the automated count with their hand count (Figure~\ref{fig:cv}). The reflection point asks whether the instrument detected the intended objects and how characteristics of the images it was given, such as contrast, resolution, or framing, affected false detections and missed objects. A confident count can still be wrong. Related observing instruments extend this work beyond still images: object tracking follows individual animals across video frames \parencite{romeroferrero2019}, and audio classification identifies species from their calls \parencite{kahl2021}.

\subsection{Analyzing and Interpreting Data}
\label{subsec:analyzing}

\begin{figure}[htbp]
\centering
\begin{illusbox}
{\centering\includegraphics[width=0.82\linewidth]{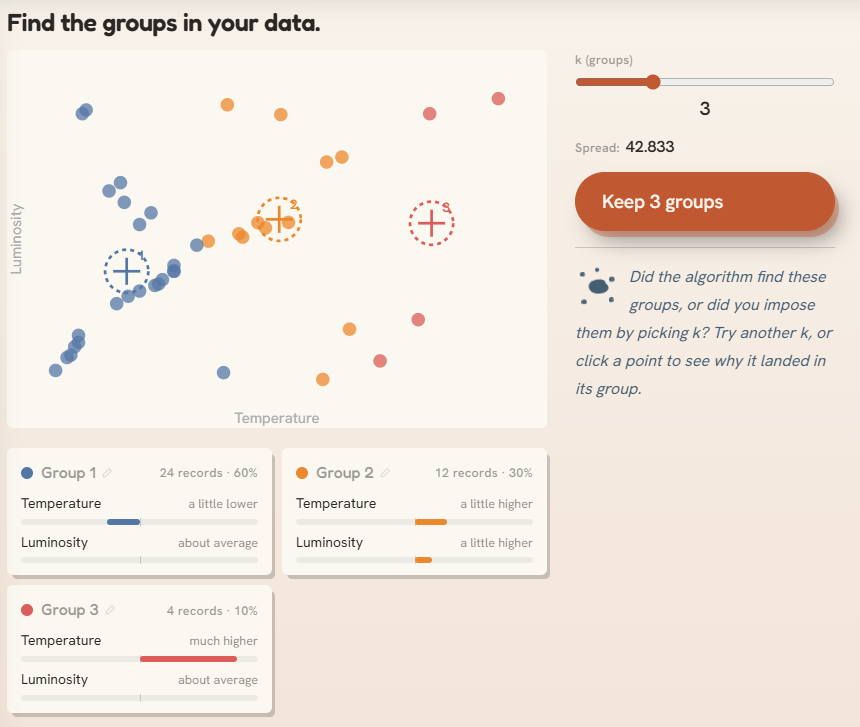}\par}
\captionof{figure}{Analyzing: Clustering}\label{fig:cluster}
{\small\RaggedRight
\textbf{What it does:} Sorts records into as many groups ($k$) as the student asks for, by similarity.\par\smallskip
\textbf{Student's task:} Set $k$, examine the resulting groups and their within-group spread, then re-run with a different $k$; click a point to inspect which feature values contributed to its assignment.\par\smallskip
\textbf{Built-in check:} ``Did the algorithm find these groups, or did you impose them by picking $k$?''---structure supported by the evidence vs.\ structure shaped by analytical choices.\par\smallskip
\textbf{Across domains:} The same AI instrument can be applied to stars, biological measurements, chemical samples, or survey responses, although the choice of variables and interpretation of the groups remain domain-specific.\par}
\end{illusbox}
\end{figure}

In data-rich fields such as astronomy and genomics, AI can identify candidate patterns in datasets too large to analyze by hand: it clusters data and fits relationships between measured quantities \parencite{jordan2015,wang2023}. Our second AI instrument is a reduced version of clustering. Clustering sorts records into as many groups as the student asks for, by similarity; its reflection point is whether those groups were discovered or imposed by that choice, which the student can test by choosing a different number and evaluating mathematical and theoretical fit (Figure~\ref{fig:cluster}). A related instrument, regression, fits a function to the data and raises a parallel reflection point: whether a curve is being forced through scatter, and whether it can be trusted beyond the range of the data. The clustering instrument explicitly fosters reflection on the AI instrument: a clustering suggested by AI is not necessarily scientifically meaningful.

\subsection{Developing and Using Models}
\label{subsec:modeling}

\begin{figure}[htbp]
\centering
\begin{illusbox}
{\centering\includegraphics[width=\linewidth]{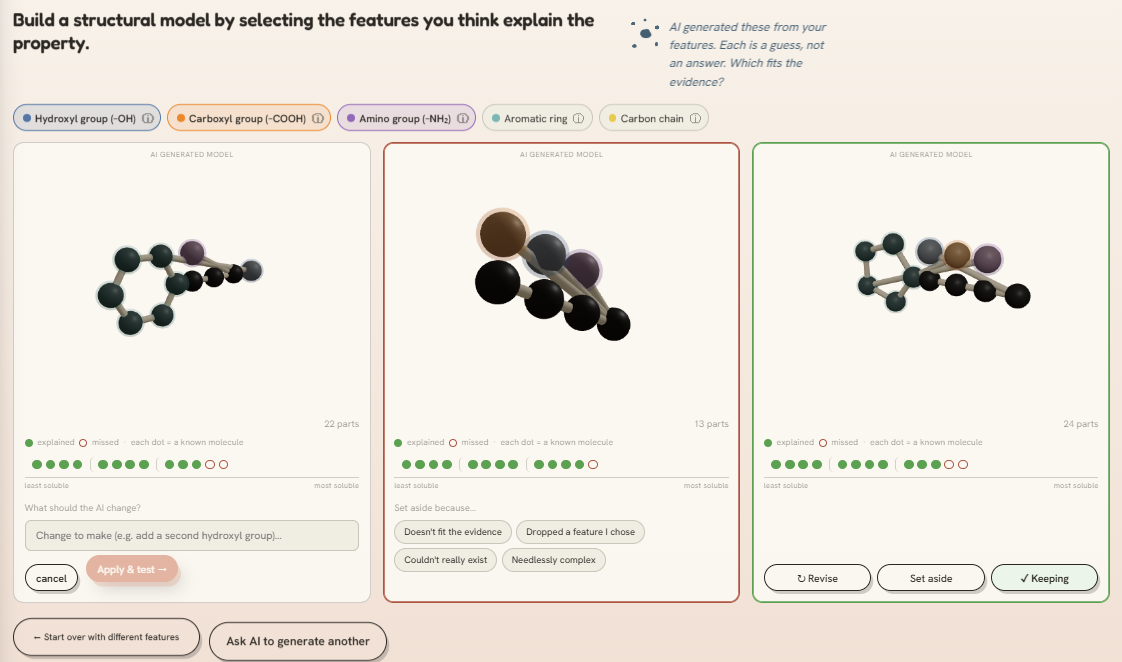}\par}
\captionof{figure}{Modeling: Generative modeler}\label{fig:genmodel}
{\small\RaggedRight
\textbf{What it does:} Turns a feature specification into candidate structures for the student to evaluate.\par\smallskip
\textbf{Student's task:} Choose the features thought to cause a property (e.g., solubility), generate candidates, and weigh each against the evidence---keep, revise, or set aside with a reason.\par\smallskip
\textbf{Built-in check:} Did it build what you specified? Is the structure scientifically plausible, and does available evidence support the intended property? The ``set aside because\ldots'' reasons make the relevant limitations explicit.\par\smallskip
\textbf{Across domains:} The same AI instrument can propose candidate structures from a specification across different domains, although the constraints and criteria used to evaluate them remain domain-specific.\par}
\end{illusbox}
\end{figure}

Generative models now design novel proteins and molecules from a target specification, producing candidate structures that need not occur in nature \parencite{watson2023}. Our third illustrative AI instrument is a pedagogically bounded version of generative design, set within the practice of developing and using models \parencite{schwarz2009}. The student can select or describe the features they think are responsible for a property, such as which structural groups make a molecule taste sweet \parencite{yun2026} or dissolve in water, and the AI instrument returns several candidate structures, each a generated structure to be evaluated. The generation itself is a ``black box,'' so the student's task lies in evaluation and critique: weighing the candidates against the evidence. Its reflection point asks whether the instrument built what the student specified, whether the structure is scientifically plausible, and whether available evidence supports the intended property (Figure~\ref{fig:genmodel}). This again is meant to foster the students' explicit reflection on AI in the scientific inquiry. Two related instruments extend the modeling toolbox more briefly: symbolic regression, which has AI propose an equation rather than only a curve and raises the reflection point of whether the formula is a genuine law or an elaborate fit to noise \parencite{schmidt2009}, and machine-learning surrogates, which give a fast approximation in place of an expensive simulation but raise the question of whether that approximation holds beyond its training range \parencite{rasp2018}. What the generative AI instrument adds is specific to this technology: aiming a model at an intended outcome and judging the result against that intent, not against how convincing it looks.

\subsection{Asking Questions and Arguing from Evidence}
\label{subsec:asking}

Two practices often precede and follow the three illustrative functions described above, and AI has a genuine role in each. At the outset, in asking questions and planning investigations, AI can survey the literature, suggest candidate hypotheses, and propose experimental designs \parencite{sourati2023,wang2023}. At the conclusion, in arguing from evidence and communicating results, it can compare findings across cases and assist in drafting figures and explanations. Each of these uses carries its own reflection point, ranging from fabricated citations at the outset to a fluent summary that obscures a consequential disagreement at the end. We note them here for completeness, but the focus of this article remains the three illustrative instruments described above, where the role of AI is most concrete and science-specific.

\subsection{Agentic AI as Collaborator}
\label{subsec:agentic}

The three AI instruments above are illustrative examples of bounded tools operated by the student within a scientific practice. Increasingly, though, (partially) autonomous agents that follow a goal across many steps of scientific inquiry are becoming more common in the sciences \parencite{gao2024}. Such agentic AI can work across an entire inquiry process rather than within a single practice: it plans the steps, calls the individual instruments in turn, reads their results, and revises its approach. Such systems already design and run experiments \parencite{boiko2023}, propose and refine their own hypotheses \parencite{googledeepmind2025}, and carry out an investigation through its full cycle with little human intervention \parencite{lu2024,szymanski2023}. What makes such agentic AI a collaborator rather than an instrument is not the range of practices it covers but the judgment it exercises within them. The researcher is left with setting the goal, directing, overseeing, and safeguarding the work, and answering for what the results mean. These systems lie beyond the single-practice AI instruments described above, and we treat agentic AI here as a horizon that instruction must prepare for.

As agentic AI becomes part of how science is practiced, students need to understand that it exists and how it works to gain a comprehensive scientific literacy \parencite{nehm2026}. Yet the autonomy that makes it useful is also what makes early reliance risky: an agentic system that can carry out a whole inquiry invites the student to hand it over completely. As it takes over large portions of the inquiry process, the student's competence might decay or never properly develop through disuse. But that competence, core scientific literacy, is exactly what is needed to supervise the system, notice its errors, and step in when it fails \parencite{bainbridge1983,parasuraman2010}.

The order of instruction, therefore, might matter. To judge what agentic AI produces and whether it can be trusted, a student needs some scientific literacy and DAIL \parencite{zhai2025}. A scientific inquiry an AI agent carries out autonomously is one that the student might have little chance to learn from. Scientific literacy can grow alongside a student's use of AI instruments, but the capacity to evaluate an agent's outputs should develop before substantial parts of an investigation are handed over. Agentic AI therefore does not replace direct observation through a microscope, nor the inquiry carried out with the individual AI instruments of Sections~\ref{subsec:observing} to \ref{subsec:asking}; it builds on them. Relied upon before sufficient scientific literacy and DAIL are in place, agentic AI risks dependence instead of understanding, leaving students who can neither judge what the AI agent returns nor acquire the scientific literacy the instruction exists to develop.

\section{Discussion}
\label{sec:discussion}

\subsection{Educational Opportunities}
\label{subsec:opportunities}

An argument for using AI instruments in science education is authenticity: pedagogically bounded instruments can preserve genuine scientific practices, such as counting populations from images, clustering complex data, and generating candidate models, while leaving interpretation and evaluation to the student. Because the AI instruments are situated within NGSS practices, the AI content does not compete with the science curriculum but extends it. Equally important, the approach treats the limitations and evaluation of the AI as an integral aspect of the instruction rather than as a hidden liability. Each AI instrument is paired with a reflection about how it can be wrong, making reasoning about its flaws an explicit part of the work, which is the core of scientific inquiry.

\subsection{Implications for AI Literacy}
\label{subsec:implications}

This design enacts discipline-based AI literacy in the sense recent work has called for \parencite{knoth2024}. Rather than presenting AI concepts like ``clustering'' in the abstract or focusing on the computational mechanics, and relying on transfer, the instruction situates each concept within an authentic AI instrument: students learn about generative modeling by specifying and evaluating generated candidates, and they learn its risks by seeing what happens when the features they choose fail to account for the evidence. The approach complements rather than replaces general frameworks such as those of \textcite{long2020} and \textcite{touretzky2019}. What accumulates across units is not a catalog of algorithms but an understanding of how AI changes the practice of science, a form of literacy that is difficult to obtain from a standalone AI literacy course but necessary for a scientifically literate public.

\subsection{Human-AI Collaboration}
\label{subsec:collaboration}

When students use the individual AI instruments, the division of work is clear. Each instrument supports a distinct scientific practice. The student sets the intent and decides whether to trust what comes back. Agentic AI, by contrast, has the capability to decide what to run next, when a result is good enough, and which line of inquiry to drop. Those are the student's decisions when the AI instruments are used by the student. With an AI agent in the loop, the line between the student's thinking and the system's ``reasoning'' becomes increasingly blurred.

Once the AI agent becomes a partly independent collaborator, validation becomes more difficult. A student can test a single clustering directly against the data, but validating a whole inquiry built from the agent's chain of choices is more demanding because each decision, intermediate output, and dependency must also be inspected. The student may come to trust the agent rather than reflect on and check it. This is where we see the contribution of the AI instruments (Sections~\ref{subsec:observing}--\ref{subsec:modeling}). Operating them is how a student learns the reflection and critical evaluation that an AI agent will demand, both to direct it and to question its outputs. Whether that critical evaluation carries over from a single set of AI instruments to a full inquiry remains to be understood. This reflects a careful, accountable partnership that practicing scientists are themselves still working out \parencite{wang2023}.

\subsection{Challenges and Limitations}
\label{subsec:challenges}

Among the challenges this vision raises, the first is over-reliance. AI instruments are fluent enough that students may take their output at face value and skip the reasoning the lesson is intended to develop. The explicit focus on reflection points is a direct response, but it is effective only insofar as the surrounding instruction prompts students to apply it. The accompanying assistant is constrained for the same reason: its responses are tied to the activity materials and the student's own work, and it redirects requests for the answer back to the evidence rather than supplying it. The reliability of each instrument also depends on the data it is given, making the selection of suitable images, variables, examples, or constraints an important part of instructional design. The second challenge is teacher preparation. Designing lessons that place AI instruments appropriately within scientific inquiry and guiding the reflection each instrument is meant to spark demands considerable expertise at a time when many teachers are still developing their own AI literacy. A third is evidential: the approach is so far a design argument, and whether operating these instruments produces the scientific and AI literacy we claim remains to be tested in classrooms. These challenges are not disqualifying, but they mark the conditions under which careful design and further research are needed.

\section{Conclusion}
\label{sec:conclusion}

AI has become a new generation of scientific instruments, distributed throughout scientific inquiry. A science education that ignores this development leaves students unprepared for science as it is now practiced. The approach we describe anchors each AI instrument to an NGSS scientific practice and pairs it with an explicit way to reflect and critically evaluate how it can mislead. It follows two aims at once: authentic scientific inquiry and DAIL that develops into a working map of where and how AI enters science. The vision is concrete because its AI instruments do real scientific work, and it is sustainable because the skepticism it cultivates is the established skepticism of science, now directed at a new class of instruments. Beyond these instruments, researchers already work with agentic AI that runs across the whole scientific inquiry. Students should understand this emerging reconfiguration of scientific inquiry and how to reflect on the AI agent's role \parencite{nehm2026}. As argued in Section~\ref{subsec:agentic}, the order of instruction matters: students should develop foundations in scientific inquiry and its AI instruments before relying on increasingly autonomous AI agents, or they risk handing the science to a system they can neither adequately check nor take over when it fails.

\section*{Declarations}
\addcontentsline{toc}{section}{Declarations}

\subsection*{CRediT authorship contribution statement}

\noindent Arne Bewersdorff: Conceptualization, Software, Writing---original draft, Writing---review \& editing, Visualization.\\
Matias Rojas: Writing---review \& editing.\\
Xiaoming Zhai: Conceptualization, Supervision, Funding acquisition, Writing---review \& editing.

\subsection*{Competing interests}

The authors declare no competing interests.

\subsection*{Funding}

The research reported here was supported by the Institute of Education Sciences, U.S. Department of Education, through Grant R305C240010 (PI Zhai). The opinions expressed are those of the authors and do not represent views of the Institute or the U.S. Department of Education.

\subsection*{Acknowledgements}

We want to thank M. Yun, K. J. Crippen, C. Xie, \& D. Bulseco for their valuable insights and pioneering work on the GenAI molecule modelling approach.

\subsection*{Use of AI in the writing process}

During the preparation of this work, the authors used ChatGPT and Anthropic Claude in order to improve the readability and language of sentences as authors are not native English speakers. After using these tools, the authors reviewed and edited the content as needed and take full responsibility for the content of the publication.

\printbibliography[title={References}]

@article{bainbridge1983,
  author  = {Bainbridge, L.},
  title   = {Ironies of automation},
  journal = {Automatica},
  year    = {1983},
  volume  = {19},
  number  = {6},
  pages   = {775--779},
  doi     = {10.1016/0005-1098(83)90046-8},
}

@article{boiko2023,
  author  = {Boiko, D. A. and MacKnight, R. and Kline, B. and Gomes, G.},
  title   = {Autonomous chemical research with large language models},
  journal = {Nature},
  year    = {2023},
  volume  = {624},
  number  = {7992},
  pages   = {570--578},
  doi     = {10.1038/s41586-023-06792-0},
}

@article{breiman2001,
  author  = {Breiman, L.},
  title   = {Statistical modeling: The two cultures},
  journal = {Statistical Science},
  year    = {2001},
  volume  = {16},
  number  = {3},
  pages   = {199--231},
  doi     = {10.1214/ss/1009213726},
}

@article{chabot2016,
  author  = {Chabot, D. and Francis, C. M.},
  title   = {Computer-automated bird detection and counts in high-resolution aerial images: A review},
  journal = {Journal of Field Ornithology},
  year    = {2016},
  volume  = {87},
  number  = {4},
  pages   = {343--359},
  doi     = {10.1111/jofo.12171},
}

@article{gao2024,
  author  = {Gao, S. and Fang, A. and Huang, Y. and Giunchiglia, V. and Noori, A. and Schwarz, J. R. and Ektefaie, Y. and Kondic, J. and Zitnik, M.},
  title   = {Empowering biomedical discovery with {AI} agents},
  journal = {Cell},
  year    = {2024},
  volume  = {187},
  number  = {22},
  pages   = {6125--6151},
  doi     = {10.1016/j.cell.2024.09.022},
}

@online{googledeepmind2025,
  author       = {{Google DeepMind}},
  title        = {Towards an {AI} co-scientist},
  date         = {2025},
  organization = {arXiv},
  url          = {https://arxiv.org/abs/2502.18864},
}

@book{hastie2009,
  author    = {Hastie, T. and Tibshirani, R. and Friedman, J.},
  title     = {The elements of statistical learning: Data mining, inference, and prediction},
  year      = {2009},
  edition   = {2},
  publisher = {Springer},
  doi       = {10.1007/978-0-387-84858-7},
}

@incollection{herdliska2024,
  author    = {Herdliska, A. and Zhai, X.},
  title     = {Artificial intelligence-based scientific inquiry},
  editor    = {Zhai, X. and Krajcik, J.},
  booktitle = {Uses of artificial intelligence in {STEM} education},
  year      = {2024},
  pages     = {179--197},
  publisher = {Oxford University Press},
  doi       = {10.1093/oso/9780198882077.003.0009},
}

@article{hmelosilver2004,
  author  = {Hmelo-Silver, C. E. and Pfeffer, M. G.},
  title   = {Comparing expert and novice understanding of a complex system from the perspective of structures, behaviors, and functions},
  journal = {Cognitive Science},
  year    = {2004},
  volume  = {28},
  number  = {1},
  pages   = {127--138},
  doi     = {10.1207/s15516709cog2801_7},
}

@article{jordan2015,
  author  = {Jordan, M. I. and Mitchell, T. M.},
  title   = {Machine learning: Trends, perspectives, and prospects},
  journal = {Science},
  year    = {2015},
  volume  = {349},
  number  = {6245},
  pages   = {255--260},
  doi     = {10.1126/science.aaa8415},
}

@article{jumper2021,
  author  = {Jumper, J. and Evans, R. and Pritzel, A. and Green, T. and Figurnov, M. and Ronneberger, O. and Tunyasuvunakool, K. and Bates, R. and {\v{Z}}{\'i}dek, A. and Potapenko, A. and Bridgland, A. and Meyer, C. and Kohl, S. A. A. and Ballard, A. J. and Cowie, A. and Romera-Paredes, B. and Nikolov, S. and Jain, R. and Adler, J. and Hassabis, D.},
  title   = {Highly accurate protein structure prediction with {AlphaFold}},
  journal = {Nature},
  year    = {2021},
  volume  = {596},
  number  = {7873},
  pages   = {583--589},
  doi     = {10.1038/s41586-021-03819-2},
}

@article{kahl2021,
  author  = {Kahl, S. and Wood, C. M. and Eibl, M. and Klinck, H.},
  title   = {{BirdNET}: A deep learning solution for avian diversity monitoring},
  journal = {Ecological Informatics},
  year    = {2021},
  volume  = {61},
  pages   = {101236},
  doi     = {10.1016/j.ecoinf.2021.101236},
}

@article{knoth2024,
  author  = {Knoth, N. and Decker, M. and Laupichler, M. C. and Pinski, M. and Buchholtz, N. and Bata, K. and Schultz, B.},
  title   = {Developing a holistic {AI} literacy assessment matrix: Bridging generic, domain-specific, and ethical competencies},
  journal = {Computers and Education Open},
  year    = {2024},
  volume  = {6},
  pages   = {100177},
  doi     = {10.1016/j.caeo.2024.100177},
}

@inproceedings{long2020,
  author    = {Long, D. and Magerko, B.},
  title     = {What is {AI} literacy? Competencies and design considerations},
  booktitle = {Proceedings of the 2020 {CHI} Conference on Human Factors in Computing Systems},
  year      = {2020},
  pages     = {1--16},
  publisher = {Association for Computing Machinery},
  doi       = {10.1145/3313831.3376727},
}

@online{lu2024,
  author       = {Lu, C. and Lu, C. and Lange, R. T. and Foerster, J. and Clune, J. and Ha, D.},
  title        = {The {AI} Scientist: Towards fully automated open-ended scientific discovery},
  date         = {2024},
  organization = {arXiv},
  url          = {https://arxiv.org/abs/2408.06292},
}

@article{machamer2000,
  author  = {Machamer, P. and Darden, L. and Craver, C. F.},
  title   = {Thinking about mechanisms},
  journal = {Philosophy of Science},
  year    = {2000},
  volume  = {67},
  number  = {1},
  pages   = {1--25},
  doi     = {10.1086/392759},
}

@book{nrc2012,
  author    = {{National Research Council}},
  title     = {A framework for {K--12} science education: Practices, crosscutting concepts, and core ideas},
  year      = {2012},
  publisher = {The National Academies Press},
  doi       = {10.17226/13165},
}

@incollection{nehm2026,
  author    = {Nehm, R. H. and Kubsch, M.},
  title     = {{AI} in contemporary scientific practice: Implications for {AI}-integrated science education},
  editor    = {Zhai, X. and Crippen, K. J.},
  booktitle = {Advancing {AI} in science education: Envisioning responsible and ethical practice},
  year      = {2026},
  volume    = {3},
  pages     = {27--48},
  publisher = {Springer},
  doi       = {10.1007/978-3-032-16871-9_2},
}

@book{ngss2013,
  author    = {{NGSS Lead States}},
  title     = {Next Generation Science Standards: For states, by states},
  year      = {2013},
  publisher = {The National Academies Press},
  doi       = {10.17226/18290},
}

@article{norouzzadeh2018,
  author  = {Norouzzadeh, M. S. and Nguyen, A. and Kosmala, M. and Swanson, A. and Palmer, M. S. and Packer, C. and Clune, J.},
  title   = {Automatically identifying, counting, and describing wild animals in camera-trap images with deep learning},
  journal = {Proceedings of the National Academy of Sciences},
  year    = {2018},
  volume  = {115},
  number  = {25},
  pages   = {E5716--E5725},
  doi     = {10.1073/pnas.1719367115},
}

@article{parasuraman2010,
  author  = {Parasuraman, R. and Manzey, D. H.},
  title   = {Complacency and bias in human use of automation: An attentional integration},
  journal = {Human Factors},
  year    = {2010},
  volume  = {52},
  number  = {3},
  pages   = {381--410},
  doi     = {10.1177/0018720810376055},
}

@article{rasp2018,
  author  = {Rasp, S. and Pritchard, M. S. and Gentine, P.},
  title   = {Deep learning to represent subgrid processes in climate models},
  journal = {Proceedings of the National Academy of Sciences},
  year    = {2018},
  volume  = {115},
  number  = {39},
  pages   = {9684--9689},
  doi     = {10.1073/pnas.1810286115},
}

@article{romeroferrero2019,
  author  = {Romero-Ferrero, F. and Bergomi, M. G. and Hinz, R. C. and Heras, F. J. H. and de Polavieja, G. G.},
  title   = {idtracker.ai: Tracking all individuals in small or large collectives of unmarked animals},
  journal = {Nature Methods},
  year    = {2019},
  volume  = {16},
  number  = {2},
  pages   = {179--182},
  doi     = {10.1038/s41592-018-0295-5},
}

@book{russell2021,
  author    = {Russell, S. and Norvig, P.},
  title     = {Artificial intelligence: A modern approach},
  year      = {2021},
  edition   = {4},
  publisher = {Pearson},
}

@article{schmidt2009,
  author  = {Schmidt, M. and Lipson, H.},
  title   = {Distilling free-form natural laws from experimental data},
  journal = {Science},
  year    = {2009},
  volume  = {324},
  number  = {5923},
  pages   = {81--85},
  doi     = {10.1126/science.1165893},
}

@article{schwarz2009,
  author  = {Schwarz, C. V. and Reiser, B. J. and Davis, E. A. and Kenyon, L. and Acher, A. and Fortus, D. and Shwartz, Y. and Hug, B. and Krajcik, J.},
  title   = {Developing a learning progression for scientific modeling: Making scientific modeling accessible and meaningful for learners},
  journal = {Journal of Research in Science Teaching},
  year    = {2009},
  volume  = {46},
  number  = {6},
  pages   = {632--654},
  doi     = {10.1002/tea.20311},
}

@article{sourati2023,
  author  = {Sourati, J. and Evans, J. A.},
  title   = {Accelerating science with human-aware artificial intelligence},
  journal = {Nature Human Behaviour},
  year    = {2023},
  volume  = {7},
  number  = {10},
  pages   = {1682--1696},
  doi     = {10.1038/s41562-023-01648-z},
}

@article{szymanski2023,
  author  = {Szymanski, N. J. and Rendy, B. and Fei, Y. and Kumar, R. E. and He, T. and Milsted, D. and McDermott, M. J. and Gallant, M. and Cubuk, E. D. and Merchant, A. and Kim, H. and Jain, A. and Bartel, C. J. and Persson, K. and Zeng, Y. and Ceder, G.},
  title   = {An autonomous laboratory for the accelerated synthesis of novel materials},
  journal = {Nature},
  year    = {2023},
  volume  = {624},
  number  = {7990},
  pages   = {86--91},
  doi     = {10.1038/s41586-023-06734-w},
}

@article{touretzky2019,
  author  = {Touretzky, D. and Gardner-McCune, C. and Martin, F. and Seehorn, D.},
  title   = {Envisioning {AI} for {K--12}: What should every child know about {AI}?},
  journal = {Proceedings of the {AAAI} Conference on Artificial Intelligence},
  year    = {2019},
  volume  = {33},
  number  = {1},
  pages   = {9795--9799},
  doi     = {10.1609/aaai.v33i01.33019795},
}

@article{wang2023,
  author  = {Wang, H. and Fu, T. and Du, Y. and Gao, W. and Huang, K. and Liu, Z. and Chandak, P. and Liu, S. and Van Katwyk, P. and Deac, A. and Anandkumar, A. and Bergen, K. and Gomes, C. P. and Ho, S. and Kohli, P. and Lasenby, J. and Leskovec, J. and Liu, T.-Y. and Manrai, A. and Zitnik, M.},
  title   = {Scientific discovery in the age of artificial intelligence},
  journal = {Nature},
  year    = {2023},
  volume  = {620},
  number  = {7972},
  pages   = {47--60},
  doi     = {10.1038/s41586-023-06221-2},
}

@article{watson2023,
  author  = {Watson, J. L. and Juergens, D. and Bennett, N. R. and Trippe, B. L. and Yim, J. and Eisenach, H. E. and Ahern, W. and Borst, A. J. and Ragotte, R. J. and Milles, L. F. and Wicky, B. I. M. and Hanikel, N. and Pellock, S. J. and Courbet, A. and Sheffler, W. and Wang, J. and Venkatesh, P. and Sappington, I. D. and V{\'a}zquez Torres, S. and Baker, D.},
  title   = {De novo design of protein structure and function with {RFdiffusion}},
  journal = {Nature},
  year    = {2023},
  volume  = {620},
  number  = {7976},
  pages   = {1089--1100},
  doi     = {10.1038/s41586-023-06415-8},
}

@article{yun2026,
  author  = {Yun, M. and Crippen, K. J. and Xie, C. and Bulseco, D.},
  title   = {From prompt engineering to modeling: Secondary science teachers' use of generative {AI} to engineer sweetener molecules},
  journal = {Chemistry Education Research and Practice},
  year    = {2026},
  doi     = {10.1039/D6RP00064A},
}

@online{zhai2025,
  author       = {Zhai, X.},
  title        = {{DAIL}: Discipline-based artificial intelligence literacy},
  date         = {2025},
  organization = {SSRN},
  doi          = {10.2139/ssrn.5745703},
}

\end{document}